\documentclass[10pt,letterpaper]{article}
\usepackage{opex3}

\begin{document}




\title{Phase dependence of electron localization in HeH$^{2+}$ dissociation with an intense few-cycle laser pulse}

\author{Kunlong Liu$^{1,3}$, Weiyi Hong$^{1,3}$, and Peixiang Lu$^{1,2}$$^*$}

\address{$^1$Wuhan National Laboratory for Optoelectronics, Huazhong University of Science and Technology,
 Wuhan 430074, China \\
 $^2$School of Science, Wuhan Institute of Technology, Wuhan 430073, China \\
 $^3$These authors contributed equally to this work}

\email{$^*$lupeixiang@mail.hust.edu.cn} 



\begin{abstract}
The electron localization in the dissociation of the asymmetric charged molecular ion HeH$^{2+}$ exposed to an intense few-cycle laser pulse is studied by solving numerically the 3D time-dependent Schr$\ddot{\mathrm{o}}$dinger equation. By varying the carrier-envelope phase (CEP) and the intensity of the pulse, the upward shift of the localization probability and the suppression of the dissociation channel He$^{2+}$+H are observed . Our analysis shows that the phenomenon is attributed to the asymmetric structure of the molecule as well as the recollision-assistant field-induced ionization of the electron wave packets localized on H$^+$ in the trailing of the pulse.
\end{abstract}


With the development of laser technology, the molecular dynamics in strong field is progressively understood in depth by scientists. Many of the processes in terms of strong-field molecular interactions have been discovered and interpreted in experimental and theoretical studies on H$_2$ and its isotopes, such as bond softening, vibrational trapping, enhanced ionization and so on \cite{Posthumus}.

Recently, the electron localization in dissociating molecules has become an interesting subject for molecular research \cite{Bandrauk2004, Roudnev,Roudnev2}. The control of electron localization in molecules on the attosecond time-scale is crucial for steering many physical and chemical processes \cite{CorkumNP, Krausz}. In a landmark experiment, Kling \textit{et al.} \cite{Kling} reported the probability to control the direction of the D$^+$ ejection (and hence the localization of the electron) in the dissociating D$_2^+$ by varying the CEP of the pulse. This constitutes the first realization of direct light-field control of a chemical reaction via the steering of electronic motion, and it has stimulated continuous interest to study the electron localization in dissociating small molecules \cite{Staudte,Tong,Grafe,He,He2,Kremer,Ray,
Sansone,Singh,Fischer,Calvert} that include two nuclei with the identical charge. More recently,
Znakovskaya \textit{et al.} \cite{Znakovskaya} have reported an experimental demonstration of electron localization in the dissociation of a heteronuclear molecule, CO, and the control of the directional emission of C$^+$ and O$^+$ fragments has been achieved in their work.

Motivated by understanding more detail of the physical mechanism for electron localization on asymmetric charged fragments, we set our efforts on theoretically studying the dissociation of a model molecular ion, HeH$^{2+}$,
in an intense few-cycle pulse.
As the simplest asymmetric charged molecule, the cation of HeH has been the subject of many works on laser-induced phenomena, such as high-order harmonic generation \cite{Lanol,Lan}, enhanced ionization \cite{Bandrauk2005}, double ionization \cite{Liao} and photodissociation \cite{Kumitriu}. The molecular system of HeH$^{2+}$ consists of three charged particles and a bound state of the system is existent \cite{Heh}.
Interacting with a strong laser field, HeH$^{2+}$ exhibits three fragmentation pathways, the dissociation channels He$^+$+H$^+$ and He$^{2+}$+H, where the electron localizes on either of the nuclei, and the ionization channel He$^{2+}$+H$^+$+e$^-$.
Hereafter, $p$ and $\alpha$ are used to denote the proton H$^+$ and alpha particle He$^{2+}$, respectively.
In the present work, the CEP effect on different dissociation channels of HeH$^{2+}$ is studied by solving numerically the 3D time-dependent Schr$\ddot{\mathrm{o}}$dinger equation for the interaction. The numerical results show an interesting phenomenon that the probability of the electron localized on $p$ shifts upwards with increasing laser intensity and the dissociation channel $\alpha$+H is suppressed for some CEPs of the pulse.
We then qualitatively explain the new findings by analyzing the electron localization during the dissociation process.

For numerical simulation we have used the three-dimensional model for HeH$^{2+}$ interacting with a linearly polarized few-cycle pulse. This model consists of one-dimensional motion of the nuclei and two-dimensional motion of the electron \cite{Roudnev,He}. Thus non-Born-Oppenheimer effects are included during the interaction.
In our model, the molecular ion is assumed to remain aligned with the linearly polarized pulse.
Rotation of the molecule is not considered
since
the molecule does not have time to rotate significantly during the dissociation (a few tens of femtoseconds in our simulation) as well as the molecule dissociates primarily along the laser polarization.
Within this model, the time-dependent Schr$\ddot{\mathrm{o}}$dinger equation can be written as ($e=m_{e}=\hbar=1$)
\begin{equation}
i\frac{\partial}{\partial t}\Psi(R,z,\rho;t)=
[H_{0}+V(t)]\Psi(R,z,\rho;t),
\end{equation}
where $H_{0}=T_{n}+T_{e}+V_{0}$ is
the field-free Hamiltonian with
\begin{eqnarray}
T_{n}=-\frac{1}{2\mu}\frac{\partial ^2}{\partial R^2},
\end{eqnarray}
\begin{eqnarray}
T_{e}=-\frac{1}{2\mu _{e}}(\frac{\partial ^2}{\partial z^2}+\frac{\partial ^2}{\partial \rho^2} +\frac{1}{\rho}\frac{\partial}{\partial \rho}),
\end{eqnarray}
\begin{eqnarray}
V_{0}=\frac{C_{p}C_{\alpha}}{\sqrt{R^2+b}}
-\frac{C_{p}}{\sqrt{(z-z_{p})^2+\rho ^2+a}}-\frac{C_{\alpha}}{\sqrt{(z-z_{\alpha})^2+\rho ^2+a}},
\end{eqnarray}
and $V(t)$ is
the electric potential including the laser-molecule interaction.
Here, $R$ is the internuclear distance, $C_{p}$ and $C_{\alpha}$ are the electric charges of the nuclei, and
$z_{p}=m_{\alpha}/(m_{p}+m_{\alpha})R$ and
$z_{\alpha}=-m_{p}/(m_{p}+m_{\alpha})R$
are the positions of $p$ and $\alpha$, respectively.
$\mu=(1/m_{p}+1/m_{\alpha})^{-1}$ and $\mu_{e}=(m_{p}
+m_{\alpha})/(m_{p}+m_{\alpha}+1)$ are the reduced masses, with $m_{p}$ and $m_{\alpha}$ the masses of $p$ and $\alpha$. The soft-core parameters $a=0.01$ and $b=0.12$ have been chosen so that the model of HeH$^{2+}$ yields the $2p\sigma$ bound state energy and equilibrium distance of -0.5317 and 3.9 a.u. \cite{Heh}, respectively. The interaction with the laser is treated in the dipole approximation and length gauge \cite{Hiskes},
\begin{eqnarray}
V(t)=[\frac{C_{\alpha}m_{p}-C_{p}m_{\alpha}}{m_{p}
+m_{\alpha}}R+(1+\frac{C_{p}+C_{\alpha}-1}{m_{p}
+m_{\alpha}+1})z]\times E(t)
\end{eqnarray}
with $E(t)=E_0 \exp[-2\ln(2)(t/\tau)^{2}]\cos (\omega t+\phi)$,
where $E_{0}$ is the peak electric field amplitude, $\tau$ is the pulse duration, $\omega$ is the central frequency, and $\phi$ is the CEP of the pulse. The electric field is chosen to be linearly polarized along the $z$ axis. Then, one can solve Eq.(1) with the Crank-Nicolson method and evaluate the time-dependent wave function.

\begin{figure}
\centering\includegraphics[width=8cm]{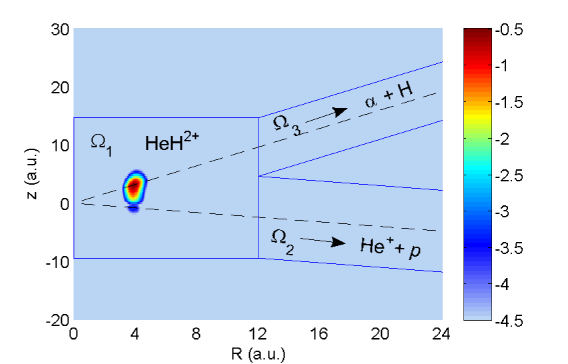}
\caption{\label{fig1} The $2p\sigma$ bound state of HeH$^{2+}$ and the schematic diagram of two different dissociation channels for HeH$^{2+}$: He$^+$+$p$ and $\alpha$+H. The up and down dash lines indicate the positions of the $p$ and $\alpha$, respectively. The color scale is logarithmic.}
\end{figure}

Before the time evolution of the wave function, an initial state is requisite. However, a stable wave function cannot be obtained directly by imaging time propagation of Eq.(1) because the ground state of HeH$^{2+}$ is unstable. This can be understood as that the vibrational wave packets are not bounded by the $1s\sigma$ curve.
In our calculation, the first excited state $2p\sigma$, which is the lowest electronic bound state of HeH$^{2+}$, is chosen to be the initial state of the system. Within the Born-Oppenheimer approximation, the initial wave function is treated as \cite{Kenfack}
\begin{eqnarray}
\Psi_{0}(R,z,\rho)=\chi_{0}(R)\Psi_{e}(z,\rho;R),
\end{eqnarray}
where $\chi_{0}(R)$ denotes the first eigenstate of the nuclei bounded by the $2p\sigma$ potential curve and $\Psi_{e}(z,\rho;R)$ is the first excited state of HeH$^{2+}$ at each fixed nuclear separation $R$. In detail, the electric wave function $\Psi_{e}(z,\rho;R)$ is found by solving the Schr$\mathrm{\ddot{o}}$dinger equation of the electronic motion at each fixed separation. The potential curve $V_{n}(R)$ of the $2p\sigma$ state can be obtained by
$V_{n}(R)=\langle\Psi_{e}(z,\rho;R)|(T_{e}+V_{0})|\Psi_{e}(z,\rho;R)\rangle$.
Then the first eigenstate of the nuclei $\chi_{0}(R)$ is solved with imaginary time propagation of the following Schr$\mathrm{\ddot{o}}$dinger equation \cite{Kenfack}
\begin{eqnarray}
i\frac{\partial}{\partial t}\chi(R,t)=[T_{n}+V_{n}(R)]\chi(R,t).
\end{eqnarray}
The wave function density distribution for the $2p\sigma$ bound state of HeH$^{2+}$ is illustrated in Fig.~\ref{fig1}.

In the present simulation, the grid ranges in $R$-direction from 0 to 25 a.u., for $z$ from -45 to 45 a.u. and for $\rho$ from 0 to 15 a.u. with 500, 300, and 50 points in the three directions, respectively. The wavelength and duration of the laser pulse are 800 nm and 5.3 fs (FWHM), respectively, while its CEP and intensity are variable.
We have adopted the approach from Ref.\cite{Roudnev} to analyze the wave function but defined a different configuration space (see Fig.~\ref{fig1}).
The two channels of dissociation are defined as
$\Omega _{2}: R>12$ and $\sqrt{(z-z_{\alpha})^2+\rho^2}<7$,
$\Omega _{3}: R>12$ and $\sqrt{(z-z_p)^2+\rho^2}<5$.
In the region $\Omega _{2}$ the electron localizes on the $\alpha$ contributing to the dissociation channel of He$^+$+$p$, whereas in $\Omega _{3}$ it localizes on the $p$ contributing to $\alpha$+H. The region describing the ionization is not contained in Fig.1. We estimate the ionization probability by storing the absorbed wave packets reaching the boundary in the $z$ direction.
The propagation of the wave function has been continued until the probability of the region $\Omega _{1}$ is converged. During the propagation, we have employed absorbing boundaries using cos$^{1/6}$-masking functions but stored the absorbed contributions as dissociation or ionization, respectively.

\begin{figure}
\centering\includegraphics[width=12.5cm]{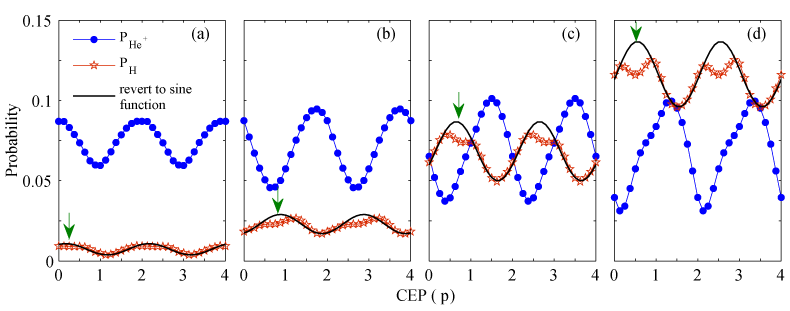}
\caption{\label{fig2} CEP dependence of the electron localization probability on the $\alpha$ (blue dots) and $p$ (red stars) for four intensities: (a) $I=4.0$, (b) $I=5.0$, (c) $I=6.0$, and (d) $I=7.0$ in units of $10^{14}$W/cm$^2$. The black thick curve indicates the reverted sine function. The green arrows indicate the suppressions of the probabilities.}
\end{figure}

Figure ~\ref{fig2} reveals the probabilities of dissociation into He$^+$+$p$ ($P_{\mathrm{He^{+}}}$) and $\alpha$+H ($P_{\mathrm{H}}$) as a function of the CEP for different pulse intensities. On the one hand, the strong CEP dependence of the dissociation channel of He$^+$+$p$ is obvious, indicating that the control of electron localization on the $\alpha$ can be achieved by adjusting the CEP of the pulse.
Moreover, the amplitude of the localization probability is larger for stronger pulse.
The probability for the dissociation limit of He$^+$+$p$, which associates with the excited electronic state, is small and increases slightly with increasing intensity. This is a result of the need to absorb more photons.
On the other hand, the CEP dependence of $P_{\mathrm{H}}$ is extraordinary.
An upward shift of $P_{\mathrm{H}}$ can be seen as the intensity increases,
and the dependence of $P_{\mathrm{H}}$ on the CEP is not a ``sine-like'' curve but it is suppressed for some CEPs, as indicated by the arrows in Fig.~\ref{fig2}. The curves of the reverted sine functions are also shown in the figure. Here,
the sine function is given by $f(x)=a_0+a_1\sin[(x+a_2)\phi]$ and the parameters $a_0$, $a_1$ and $a_2$ are fitted with the values of $P_\mathrm{H}$. To quantify the suppression, we define the absolute suppressed probability as the maximal deviation of the dissociation probability from the reverted sine function.
The suppressed probabilities are 0.003, 0.005, 0.012 and 0.02 for Fig.~\ref{fig2}(a)--(d), respectively, which means that the suppression is deeper for higher laser intensity.

\begin{figure}
\centering\includegraphics[width=9cm]{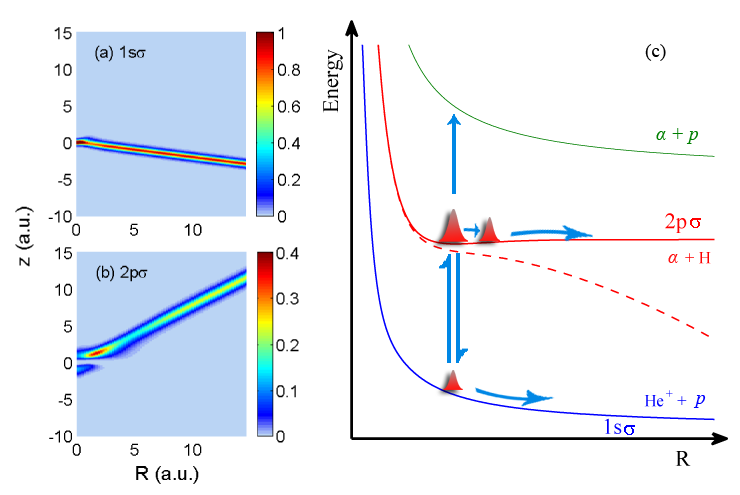}
\caption{\label{fig3} The electron density distributions of (a) the ground state and (b) the first excited state of HeH$^{2+}$ at each fixed nuclear distance. The $\rho$-dimension has been integrated. (c) A simple model for the dissociation of HeH$^{2+}$. The dashed line illustrates the dressed $2p\sigma$ state of HeH$^{2+}$ in a static field, and the Gaussians illustrate the wave packets. The color scales in (a) and (b) are linear.}
\end{figure}

In order to understand the processes of the
different dissociation channels during the interaction,
we illustrate the model for the dissociation of HeH$^{2+}$.
First of all, the electron density distributions of the lowest two states of HeH$^{2+}$ at each fixed nuclear distance
are depicted in Fig.~\ref{fig3}(a) and (b), respectively.
One can see that the electron density concentrates at the $\alpha$ in the $1s\sigma$ state while it concentrates at the $p$ in the $2p\sigma$ state. Thus the localization of the electron is related to the energy level of the molecule. If the HeH$^{2+}$ dissociates following the $1s\sigma$ potential curve, the electron would localize on the $\alpha$ contributing to the He$^+$+$p$ channel. Similarly, if the HeH$^{2+}$ dissociates following the $2p\sigma$ potential curve, the electron would localize on the $p$ contributing to the $\alpha$+H channel.
Then we illustrate the dissociation process of HeH$^{2+}$ in Fig.~\ref{fig3}(c). First, the wave packets are bounded in the $2p\sigma$ state.
When the laser pulse is applied, the electric field dresses down the $2p\sigma$ potential curve \cite{Bandrauk2005} creating vibrational wave packets. Then the populations are transferred between the $2p\sigma$ and $1s\sigma$ states by the external field during the interaction.
Part of the wave packets are ionized when the maximum light field is around.
With Fig.~\ref{fig3}(a) and (b) in mind,
the dissociating wave packets along the $1s\sigma$ and $2p\sigma$ curves result in the electron localization on the $\alpha$ and $p$ (and hence the dissociation channels of He$^+$+$p$ and $\alpha$+H), respectively.

Based on the dissociation model of HeH$^{2+}$,
the simulation results in Fig.~\ref{fig2} can be qualitatively explained as follows.
On the one hand, the electron motion is sensitive to the CEP of the few-cycle laser pulse. When the CEP varies, the population transfer between the $1s\sigma$ and $2p\sigma$ states during the interaction will be modulated, resulting in the CEP-dependent probabilities of the two dissociation channels.
On the other hand, according to Ref.\cite{Bandrauk2005}, the $2p\sigma$ curve is dressed down in both cases of $E>0$ and $E<0$. Therefore, the $2p\sigma$ state of HeH$^{2+}$ is always dressed down as the field oscillates.
The $2p\sigma$ curve would be dressed even lower if the intensity of the pulse is higher \cite{Bandrauk2005}, leading to more dissociative wave packets on the $2p\sigma$ potential curve.
With Fig.~\ref{fig3}(b) in mind, this means that more electron density concentrates at the $p$ in the dissociation process and thus the probability of dissociation channel $\alpha$+H shifts upwards when the pulse intensity increases.
In contrast, the electron localization probability on the $\alpha$ does not shift upwards because of the asymmetric distribution of the wave packets in the initial state. In the initial bound state (see Fig.1), the wave packets are primarily localized on the $p$. Thus the wave packets which localize on the fragment $\alpha$ during the dissociation have to be transferred from the $p$ by the external field. However, due to the oscillation of the pulse, most of the wave packets localized on the $\alpha$ are driven back to the $p$ if the electric field points to the $-z$ direction, as shown in Fig.4. If the intensity increases, more wave packets will be transferred from the $p$ to
the $\alpha$ when the field points to the $+z$ direction. But meanwhile, the wave packets driven from the $\alpha$ back to the $p$ will also increase when the field reverses the direction. Until the field becomes weak, a small part of the wave packets are left around the $\alpha$ because of the deep potential well and subsequently dissociate with the $\alpha$. Therefore, the localization probability $P_{\mathrm{He}^+}$ predominately depends on the potential well of the $\alpha$ but it does not increase significantly with the intensity.


\begin{figure}
\centering\includegraphics[width=10cm]{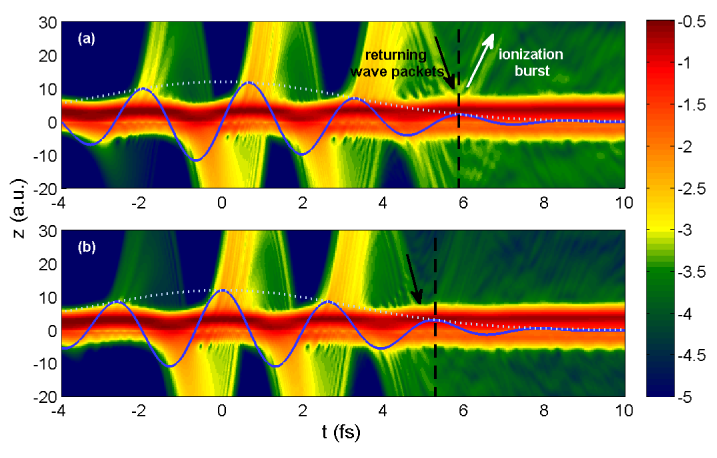}
\caption{\label{fig4} Time evolution of the wave function density integrated over the $\rho$- and $R$-dimensions. The intensity of the pulse is $7\times10^{14}\mathrm{W/cm}^2$, and the CEPs are $0.5\pi$ and $\pi$ for (a) and (b), respectively. The solid and dotted lines illustrate the electric force on the electron and the profile of the electric field, respectively. The dashed lines indicate the different times when the field reaches its peak value. The color scale is logarithmic. }
\end{figure}

\begin{figure}
\centering\includegraphics[width=10cm]{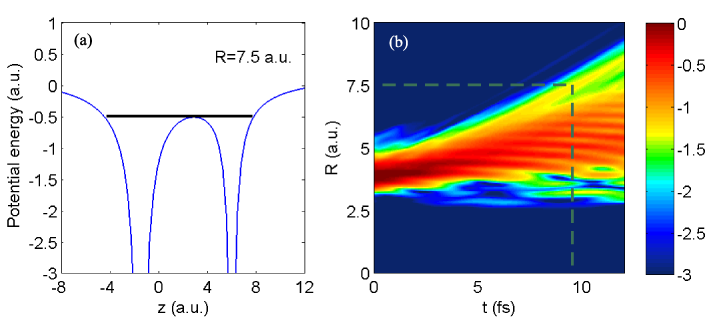}
\caption{\label{fig5} (a) The interatomic barrier potential. (b) the time evolution of the probability density for R-dimension. The laser parameters are the same as those in Fig.4(a). The thick line in (a) indicates the energy of $2p\sigma$ state at $R_0=7.5$ a.u.. The dash line in (b) indicates the time when the dissociative wave packets reach $R_0$. The color scale is logarithmic.}
\end{figure}

Next, we focus on the mechanism responsible for the suppression of the dissociation probability of $\alpha$+H. To gain further insight into the interaction between HeH$^{2+}$ and the laser pulse, the temporal evolution of the electron density is depicted in Fig.~\ref{fig4}.
The pulse intensity is $7\times10^{14}\mathrm{W/cm}^2$, and the CEPs are $0.5\pi$ and $\pi$ for Fig.~\ref{fig4}(a) and Fig.~\ref{fig4}(b), respectively.
The solid curve indicates the electric force $F(t)=-E(t)$ on the electron.
As shown in the picture, after the laser pulse is applied, the bound electron density is partially released when the field reaches the vicinity of its maximum. One part of the released electron wave packets are driven away and ionized. The other part of them return to the nuclei as the electric field changes its direction \cite{Corkum,cao,zhang,hong} and rescatter with the bound wave packets.
Then we show the interatomic barrier potential and the time evolution of the probability density for R-dimension in Fig.5.
In the present model of HeH$^{2+}$, the height of the barrier crosses the energy of the $2p\sigma$ state at $R_0=7.5$ a.u., as shown in Fig.~\ref{fig5}(a). According to Fig.~\ref{fig5}(b), the time for the dissociative wave packets to reach $R_0$ is at about 9.5 fs.
After $t=9.5$ fs, the electric field becomes very weak and the wave packets transfer between the nuclei is blocked by the increasing interatomic barrier. Then the molecule dissociates with the electron localizing on one of the two dissociating nuclei. This is consistent with our dissociation model of HeH$^{2+}$.

For the CEP of $0.5\pi$, as shown in Fig.~\ref{fig4}(a),
the returning wave packets at about $t=6$ fs interfere with the bound wave packets on the $p$ and then the density distribution is modulated. One can see that there are bumps in the bound wave packets, indicated by the black arrow, and we deduced that part of the bound wave packets are excited to higher states.
At the same time, the electric field just reaches its peak value. Although the instantaneous strength of the field is weak, part of the wave packets localized on the $p$ after excitation are still ionized by the field, leading to the week ionization burst marked in the figure.
As a result, the probability of the electron localization on the $p$ is suppressed. In the case that the CEP equals $\pi$ in Fig.~\ref{fig4}(b), the returning time of the wave packets is not the exact time at which the electric field reaches its peak value.
Although the electric field at $t=5.3$ fs in (b) is stronger than that at $t=6$ fs in (a), the ionization burst is weaker at $t=5.3$ fs in (b).
Therefore, the suppression of the localization probability does not happen for all CEPs.

On the other hand, the deepest suppressions of the localization probabilities appear to be at different CEPs for different intensities in Fig.2. It has been shown that the dynamics of the electronic motion in molecules is affected by both the external laser field and the diffraction effects \cite{He2}. In Fig.2, the regions for the suppression are mainly around 0.5 $\pi$ for different intensities, which are determined by the external field. However, due to the diffraction effects, the electronic motion will be modulated for different intensities. Therefore, the positions for the deepest suppression are changed for different intensities.
In addition,
for more intense laser pulse, the electric field is stronger and the returning wave packets take more kinetic energy, and thus more electron density will be ionized from the dissociating $p$. This comes to the result that the suppression of localization probability is deeper for higher intensity in Fig.~\ref{fig2}.
One may expect that such ionization burst would appear at the other side if the electric field is reversed.
However, the potential well of the $\alpha$ is deep and the energy level of the $1s\sigma$ state is very low. The electron firmly localizes on the $\alpha$ during the dissociation.
In contrast, the energy level of the $2p\sigma$ state is relatively higher and the potential well of the $p$ is shallow.
The electron localized on the $p$ would be easily ionized if it is excited by the returning wave packets.
Therefore, the suppression occurred in the dissociation channel of $\alpha$+H rather than He$^+$+$p$.

Finally, we note that the CEP dependence of the electron localization in the asymmetric charged molecular system of HeH$^{2+}$ is different from that in the previous studies on HD$^+$ \cite{Roudnev,Roudnev2}, in which the upward shift and suppression of the dissociation probabilities have not been reported. Moreover, the electron localizations on the $p$ and $\alpha$ are different by changing $\pi$ of the CEP. This is because the mechanism responsible for the electron localization in HeH$^{2+}$ dissociation is different from that in HD$^{+}$. For HD$^{+}$, the asymmetric localization on either the $p$ or $d$ (D$^+$) is induced by the coherent superposition of dissociative wave packets on the gerade and the ungerade states, i.e., the $1s\sigma_{g}$ and $2p\sigma_{u}$ states.
However, the localization in HeH$^{2+}$ dissociation is the result of the dissociative wave packets on the lowest two states which have asymmetric density distributions.
Compared to another previous experimental work on the dissociation of CO$^+$ \cite{Znakovskaya}, the asymmetries of both the dissociation channels, He$^+$+$p$ and C$^+$+O, where the electron is localized on the larger nucleus, shows the similar feature of the CEP dependence. There is no comparison for the other channel since the intensity dependence of it in asymmetric molecule has not been reported before.


In conclusion, we have studied the electron localization in the dissociation of the asymmetric molecular ion HeH$^{2+}$ by solving numerically the 3D time-dependent Schr$\ddot{\mathrm{o}}$dinger equation. The upward shift of the localization probability and the suppression of the dissociation have been observed. The physical mechanism responsible for the suppression is that the electron wave packets which have already localized on the dissociating nucleus is excited by the returning wave packets and subsequently ionized by the field at the trailing of the pulse.
The suppression of dissociation is also related to the localization of the electron as well as its corresponding energy level.
Generally, the phenomenon of the suppression occurs in the dissociating asymmetric molecular ions where the electron at higher energy state is localized on the nucleus with lower potential well, such as HeH$^{2+}$ and LiH$^{3+}$.
Though the asymmetric molecular ion HeH$^{2+}$ in the present work is a model system, the electron localization on the larger nucleus shows a similar CEP dependence as that in CO$^+$.

This work was supported by the National Natural Science Foundation
of China under Grants No. 60925021, 10734080 and the National
Basic Research Program of China under Grant No. 2011CB808103. This work was partially supported by the State Key Laboratory of Precision Spectroscopy of East China Normal University.
\end{document}